\documentclass{article}
\usepackage{arxiv}
\usepackage[utf8]{inputenc} 
\usepackage[T1]{fontenc}    
\usepackage{hyperref}       
\usepackage{url}            
\usepackage{booktabs}       
\usepackage{amsfonts}       
\usepackage{nicefrac}       
\usepackage{microtype}      
\usepackage[none]{hyphenat}
\usepackage{lipsum}
\usepackage{amsmath}
\usepackage{multirow}
\usepackage{stfloats}
\usepackage{graphicx}
\usepackage{booktabs}
\usepackage{array}
\usepackage{float}
\usepackage{subcaption}
\usepackage{comment}
\usepackage{xurl} 
\usepackage{hyperref}
\hypersetup{breaklinks=true}
\usepackage{siunitx}    

\title{Explainable Multi-Modal Deep Learning for Automatic Detection of Lung Diseases from Respiratory Audio Signals}

\author{
S~M~Asiful~Islam~Saky$^{1}$, Md~Rashidul~Islam$^{1}$, 
Md~Saiful~Arefin$^{1}$, Shahaba~Alam$^{1}$\\[3pt]
$^{1}$Albukhary International University, Malaysia\\
\texttt{saky.aiu22@gmail.com, mdislam0996@gmail.com,}\\
\texttt{mohammadsaifularefin@gmail.com, shahalam7127@gmail.com}
}

\begin{document}
\maketitle

\begin{abstract}
Respiratory diseases remain major global health challenges, and traditional auscultation is often limited by subjectivity, environmental noise, and inter-clinician variability. This study presents an explainable multimodal deep learning framework for automatic lung-disease detection using respiratory audio signals. The proposed system integrates two complementary representations, such as a spectral-temporal encoder based on a CNN–BiLSTM Attention architecture, and a handcrafted acoustic-feature encoder capturing physiologically meaningful descriptors such as MFCCs, spectral centroid, spectral bandwidth, and zero-crossing rate. These branches are combined through late-stage fusion to leverage both data-driven learning and domain-informed acoustic cues. The model is trained and evaluated on the Asthma Detection Dataset Version 2 using rigorous preprocessing, including resampling, normalization, noise filtering, data augmentation, and patient-level stratified partitioning. The study achieved strong generalization with 91.21\% accuracy, 0.899 macro F1-score, and 0.9866 macro ROC–AUC, outperforming all ablated variants. An ablation study confirms the importance of temporal modeling, attention mechanisms, and multimodal fusion. The framework incorporates Grad-CAM, Integrated Gradients, and SHAP, generating interpretable spectral, temporal, and feature-level explanations aligned with known acoustic biomarkers to build clinical transparency. The findings demonstrate the framework’s potential for telemedicine, point-of-care diagnostics, and real-world respiratory screening.
\end{abstract}


\section{Introduction}
\label{sec:intro}

Respiratory diseases such as Asthma, Chronic Obstructive Pulmonary Disease (COPD), and pneumonia constitute a significant global health issue, markedly impacting morbidity and mortality rates worldwide. Post-2020 reports from the World Health Organization (WHO) indicate that chronic respiratory disorders rank among the primary causes of mortality and disability, affecting more than 80 million diagnosed individuals and many more who remain undiagnosed \cite{who2025respiratory}. This underscores the critical importance of early and accurate identification of these conditions, which is essential not only for enabling timely clinical intervention but also for alleviating the significant socioeconomic strain these diseases impose on global healthcare systems.

Historically, pulmonary auscultation listening to lung sounds via a stethoscope has been fundamental to respiratory diagnosis. Despite being non-invasive and clinically valuable, traditional auscultation is plagued by intrinsic subjectivity and a substantial reliance on the clinician's auditory proficiency \cite{sabry2024lung}. Diagnostic inconsistencies frequently occur due to environmental interference, the transient nature of atypical sounds, and inter-observer variability, especially among less experienced clinicians. As a result, the diagnostic process may overlook subtle acoustic indicators of early disease onset \cite{sabry2024lung}. These limitations highlight the need for automated, objective, and noise-resilient diagnostic systems capable of analyzing respiratory sounds with clinical consistency, thereby reducing human subjectivity and improving early disease detection.

The advent of computer-aided auscultation, propelled by advancements in signal processing and artificial intelligence (AI), has begun to transform this diagnostic framework. Deep learning offers the ability to autonomously extract and learn hierarchical representations from intricate biomedical audio signals. Since 2020, various studies have utilized deep architectures specifically Convolutional Neural Networks (CNNs) for spectral feature extraction and Recurrent Neural Networks (RNNs), including Long Short-Term Memory (LSTM) and Bidirectional Gated Recurrent Units (BiGRUs), for modeling temporal dependencies to classify respiratory sounds \cite{joy2021classification, kim2021respiratory}. Hybrid CNN-LSTM frameworks have improved accuracy by integrating spatial and temporal feature modeling \cite{alqudah2022deep, yadav2024automated}, while attention mechanisms have enhanced interpretability by highlighting diagnostically pertinent acoustic segments \cite{gopi2025deep, sanjana2023attention}. Handcrafted acoustic descriptors capture physiologically interpretable cues such as airflow irregularities, whereas deep spectral–temporal models excel at nonlinear pattern recognition. Their combination has the potential to bridge human-interpretable reasoning with data-driven feature discovery—an area insufficiently explored in existing literature.

Despite these advancements, two significant limitations remain in prior studies. First, most approaches predominantly depend on either deep-learned embeddings such as  mel-spectrograms or handcrafted acoustic features such as MFCCs, spectral centroid, Zero-Crossing Rate , but rarely establish a cohesive multimodal framework that integrates both data-driven and domain-informed representations. This constraint impairs robustness under real-world acoustic variability \cite{karaarslan2024respiratory}. Second, the interpretability of deep learning models remains limited; the black-box nature of these systems hinders clinical trust, transparency, and regulatory acceptance. Although high accuracy is often reported, the decision-making processes remain unclear to clinicians. The explicit incorporation of Explainable AI (XAI) methodologies including Grad-CAM, Integrated Gradients, and SHAP is still inadequately explored in respiratory sound analysis, despite their capacity to elucidate model reasoning and relate it to physiological phenomena \cite{shehab2024deep}. Explainability is increasingly mandated for medical AI systems, as regulatory frameworks and clinicians require transparent, justifiable decisions, making the integration of XAI not optional but essential for real-world deployment.

This study proposes an explainable hybrid deep learning framework for automated multiclass lung-disease detection from respiratory sounds. The architecture integrates a CNN-BiLSTM-Attention network for deep spectral–temporal representation learning together with a parallel handcrafted-feature encoder that captures statistically interpretable acoustic cues. These complementary streams are fused at the representation level to enhance diagnostic robustness and generalization across diverse respiratory conditions. To ensure transparency, the framework incorporates a comprehensive XAI strategy employing Grad-CAM, Integrated Gradients, and SHAP to provide spatial, spectral, and feature-level interpretability aligned with clinically recognized auscultation patterns. Overall, the system is designed to deliver both high diagnostic accuracy and clinically meaningful explanations across five target categories: Asthma, COPD, Bronchial, Pneumonia, and Healthy.

The remainder of this paper is organized as follows. Section~\ref{sec:litreview} reviews recent advances in deep learning, multimodal fusion, temporal modeling, and XAI for respiratory sound analysis. Section~\ref{sec:methodology} details the proposed framework, including preprocessing, feature extraction, hybrid architecture design, training strategy, and evaluation protocol. Section~\ref{sec:results_discussion} presents the experimental results, including learning curves, classification performance, and interpretability analysis. Finally, Section~\ref{sec:conclusion} concludes the study and outlines potential directions for future work.

\section{Literature Review}
\label{sec:litreview}

\subsection{Deep Learning in Respiratory Sound Classification}

The rapid evolution of deep learning has transformed the landscape of respiratory sound analysis, establishing computer-aided auscultation as a cornerstone of modern pulmonary diagnostics. Diseases such as asthma, COPD, bronchitis, and pneumonia manifest through complex acoustic patterns—wheezes, crackles, and rhonchi—that are often difficult to reliably discern using traditional auscultation due to its inherent subjectivity and dependence on clinician expertise. Deep learning models offer an objective and scalable alternative, capable of capturing subtle spectro–temporal variations in lung sounds with enhanced consistency and precision.

Since 2020, numerous studies have explored CNNs and RNNs, particularly LSTM-based architectures, for automatic respiratory sound classification. These models typically employ preprocessing techniques such as Mel-Frequency Cepstral Coefficients (MFCCs) or mel-spectrogram transformations to represent raw audio in formats that preserve both spectral and temporal information. Joy and Haider \cite{joy2021classification} utilized CNNs trained on MFCC features to classify respiratory disorders across time–frequency domains, outperforming traditional machine learning baselines. Similarly, Kim et al.\ \cite{kim2021respiratory} applied CNNs with pretrained feature extractors to differentiate abnormal respiratory sounds, specifically identifying crackles and wheezes with high accuracy. Vasava and Joshiara \cite{vasava2023different} extended this paradigm by converting audio recordings into 2D spectrograms and applying deep convolutional networks, demonstrating the effectiveness of CNNs in extracting discriminative diagnostic cues. Hu et al.\ (2021) further improved CNN robustness by incorporating depthwise separable convolutions and attention pooling \cite{hu2021underwater}. The feasibility of deep learning approaches for lung disease recognition using acoustic signals has also been reinforced by recent systematic reviews \cite{sfayyih2023review}.

Subsequent research continued refining CNN-based frameworks. Demir et al.\ \cite{demir2019convolutional} combined pretrained CNNs with traditional classifiers to analyze spectrograms, while Roy and Satija introduced RDLINet, a lightweight Inception-based architecture designed for efficient respiratory disease detection \cite{roy2023rdlinet}. Yadav et al.\ \cite{yadav2024automated} proposed a CNN-driven classifier that achieved strong generalization across pathological and non-pathological categories, whereas Wang et al.\ (2024) developed a CNN ensemble with residual attention modules to capture complex acoustic variations in pediatric wheeze detection \cite{wang2024towards}. Additionally, specialized CNN-based systems have been applied for COPD-specific sound analysis, demonstrating high discriminative capability \cite{srivastava2021deep}. Collectively, these works establish CNNs as a reliable foundation for scalable respiratory sound interpretation.

\subsection{Hybrid and Multimodal Feature Fusion Approaches}

While CNNs effectively capture localized spectral patterns, they often fall short in modeling long-term temporal dependencies intrinsic to breathing cycles. To address this limitation, hybrid models combining CNNs with recurrent networks such as LSTMs have been increasingly adopted. Yadav et al.\ \cite{yadav2024automated} introduced a hybrid CNN–LSTM framework that leveraged CNNs for feature extraction and LSTMs for sequential modeling, achieving superior generalization across multiple lung conditions. Similarly, Alqudah et al.\ \cite{alqudah2022deep} evaluated several deep architectures and demonstrated that CNN–LSTM integration enhances temporal representation learning and resilience to noise in raw respiratory recordings. Building on this, Petmezas et al.\ (2022) proposed multi-branch CNN–BiLSTM hybrids that incorporated channel attention and feature gating for improved interpretability and cross-patient generalization \cite{petmezas2022automated, abhishek2024multimodal, zhao2022automatic}. Furthermore, dual-channel signal processing pipelines have been explored to jointly exploit complementary acoustic representations, enabling richer and more discriminative fusion spaces \cite{zhang2024research}.

Beyond purely deep architectures, recent efforts have focused on multimodal feature fusion, integrating handcrafted acoustic descriptors with deep-learned embeddings to exploit their complementary strengths. Karaarslan et al.\ \cite{karaarslan2024respiratory} adopted a two-stage approach that extracted statistical sound properties to form numerical feature vectors, subsequently refined through feature selection before classification. Gopi and Varghese \cite{gopi2025deep} proposed a multi-input architecture combining CNN, ResNet, and attention mechanisms, each operating on distinct feature representations such as MFCCs, Chroma STFT, and mel-spectrograms before fusion. Similarly, Shehab et al.\ \cite{shehab2024deep} and Singh et al.\ \cite{singh2025evaluation} investigated deep feature-level fusion across multiple pretrained CNN encoders to strengthen discriminative performance under variable acoustic conditions. Recent multimodal frameworks by Zhang et al.\ (2024) further emphasized cross-modal alignment strategies, integrating audio–text embeddings and self-attention-based fusion to correlate model predictions with clinically relevant descriptors \cite{zhang2024research, ohmshankar2025deep}. Additionally, multi-channel expert models such as CNN-MoE have demonstrated that channel-specific specialization can further enhance robustness when classifying heterogeneous respiratory anomalies \cite{pham2020robust}. Despite these advancements, most existing fusion systems remain constrained to either early (feature-level) or late (decision-level) fusion, while comprehensive end-to-end deep fusion pipelines—capable of jointly optimizing both handcrafted and learned features—remain sparsely explored in the literature.

\subsection{Attention Mechanisms and Temporal Modeling}

Temporal modeling plays a crucial role in respiratory sound analysis since pathological events such as crackles or wheezes occur intermittently within the breathing cycle. To capture these dynamics, attention mechanisms and bidirectional recurrent structures have gained increasing prominence. Chamberlain et al. \cite{chamberlain2016application} developed a lightweight CNN--BiGRU model with skip connections, combining spectral and temporal processing while maintaining computational efficiency. Pham et al. \cite{pham2020robust} proposed a deep learning framework incorporating various spectrogram representations and back-end recurrent networks to classify respiratory anomalies. Similarly, Perna and Tagarelli \cite{perna2019deep} demonstrated that RNN-based architectures outperform conventional models by effectively capturing both anomaly-driven and pathology-driven sound variations. For telemedicine applications, robust deep learning classification of respiratory sounds emphasizes temporal modeling for enhanced reliability [\cite{lo2022explainable}]. More recently, Sanjana et al. \cite{sanjana2023attention} introduced attention-based CRNN architectures that dynamically emphasized diagnostically relevant sound segments, yielding substantial improvements in multiclass respiratory disease identification. Advanced architectures, such as Self-Attention based Hybrid CNN-LSTM, have been introduced specifically to improve classification accuracy for respiratory sounds [\cite{bhushan2024self}]. Additionally, Li et al. (2025) presented multi-head self-attention and temporal gating modules to localize acoustic anomalies within continuous recordings, improving interpretability and diagnostic alignment \cite{li2025tm2sp}. Other robust and interpretable temporal models, such as Temporal Convolutional Networks, are also highly effective for event detection in lung sound recordings [\cite{fernando2022robust}]. These advancements highlight the growing importance of attention-enhanced temporal modeling in capturing the non-stationary nature of respiratory signals.

\subsection{Explainable AI (XAI) in Medical Audio Diagnosis}

Despite the strong performance of deep learning in medical audio analysis, the opaque “black-box’’ nature of these models continues to hinder clinical adoption. Explainable AI (XAI) has therefore become essential for improving transparency, interpretability, and trust in automated diagnostic systems. Techniques such as Grad-CAM, Integrated Gradients, and SHAP provide insight into model behavior by highlighting the spectral regions that most strongly influence classification outcomes \cite{akman2024audio, saky2025enhanced}. Recent studies in respiratory acoustics further demonstrate that XAI can align model attention with clinically relevant patterns such as wheezes and crackles, enhancing diagnostic interpretability \cite{lo2022explainable}. Broader analyses of XAI in healthcare emphasize the necessity for transparent and accountable AI pipelines, particularly in high-risk clinical decision-making \cite{saraswat2022explainable, amann2020explainability}. Complementary systematic reviews in respiratory audio diagnosis reinforce the critical role of explainability for clinical reliability and deployment \cite{kapetanidis2024respiratory}.

Most existing works in respiratory sound analysis predominantly rely on post-hoc explanations without intrinsic interpretability, and often utilize either handcrafted acoustic descriptors or deep spectral-temporal representations in isolation, compromising robustness. While hybrid CNN-RNN and attention-based systems enhance modeling, their decision processes remain opaque and fail to integrate domain-informed features with learned representations in parallel. This underscores the need for a unified, multimodal, and explainable hybrid framework—such as the proposed CNN-BiLSTM-Attention network complemented by handcrafted acoustic cues and a comprehensive XAI pipeline—which this work addresses.

\section{Methodology}
\label{sec:methodology}

\subsection{Overview of the Proposed Framework}

\begin{figure}[h]
    \centering
    \includegraphics[width=16cm]{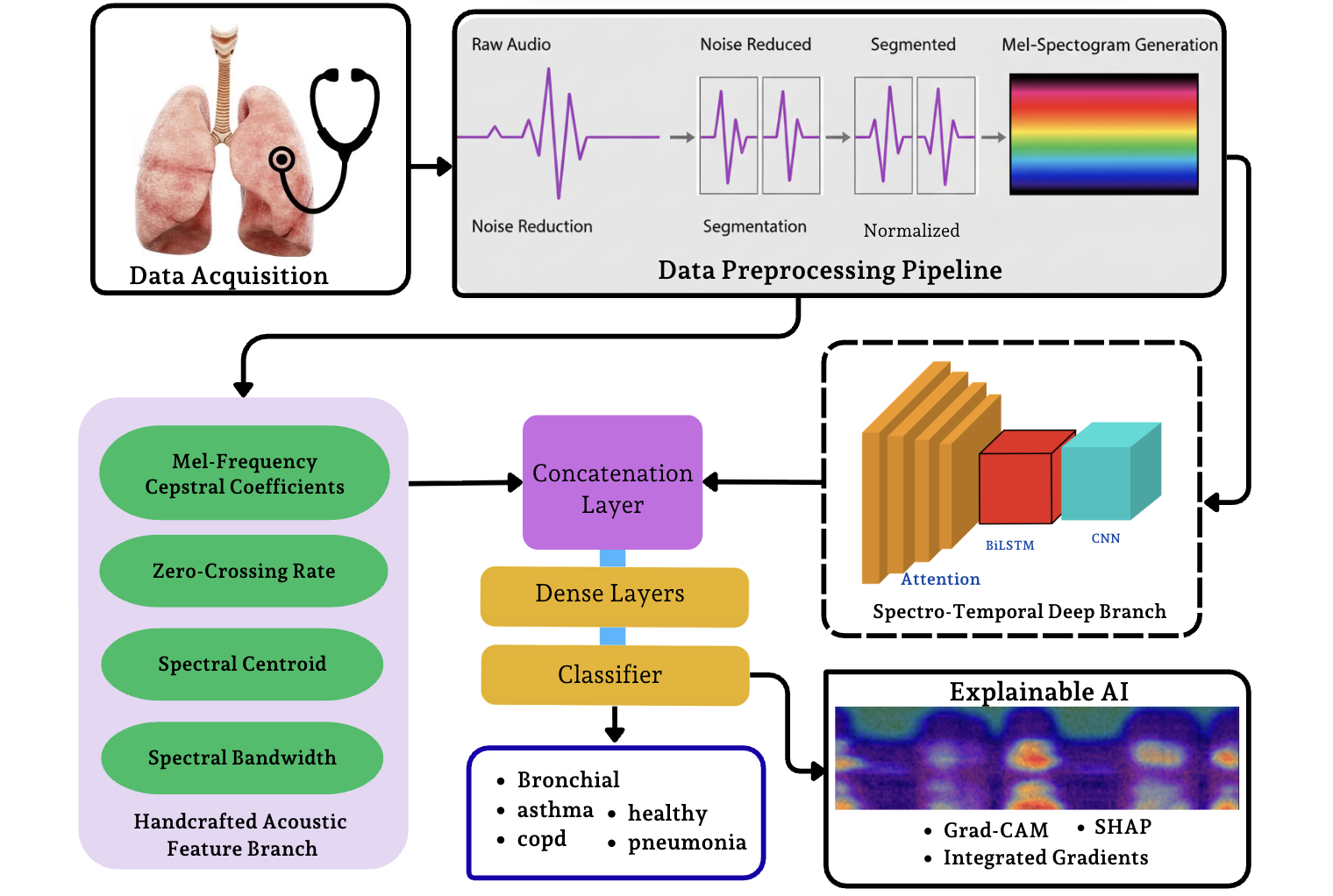}
    \caption{Proposed Hybrid Multimodal CNN–BiLSTM–Attention Framework with integrated XAI}
    \label{fig:methodology}
\end{figure}

The proposed framework, illustrated in Fig.~\ref{fig:methodology}, consists of a multimodal deep learning pipeline for automated lung disease detection from respiratory audio signals. The system begins with a standardized preprocessing stage involving resampling, normalization, temporal trimming or padding, and noise-robust augmentations to ensure acoustic consistency across recordings. Feature extraction is then performed through two parallel branches: a deep-learning branch that converts audio into mel-spectrograms and processes them using a CNN-BiLSTM Attention architecture to capture localized spectral cues, temporal dynamics, and diagnostically salient sound segments; and a handcrafted branch that extracts interpretable acoustic features such as MFCCs, spectral centroid, spectral bandwidth, Zero-Crossing Rate, and chroma features, which are encoded via dense layers. The representations from both branches are fused using a late-fusion strategy and passed to a softmax classifier to categorize samples into Asthma, COPD, Bronchial, Pneumonia, or Healthy. To ensure transparency and clinical trust, the framework incorporates Grad-CAM, Integrated Gradients, and SHAP, providing complementary spectral, temporal, and feature-level explanations of the model's decisions.

\subsection{Data Acquisition}

The experimental analysis in this study was conducted using the publicly available Asthma Detection Dataset Version~2 \cite{tawfik2022asthma}, a curated collection of respiratory sound recordings designed for diagnostic classification tasks. The dataset encompasses five clinically significant respiratory conditions: Asthma, Bronchial, COPD, Healthy, and Pneumonia. A detailed class-wise distribution of the 1,211 audio samples, including their respective percentages, is provided in Table~\ref{tab:dataset_distribution}. Each class includes a diverse set of patient recordings captured under varying environmental and acoustic conditions, ensuring heterogeneity in terms of breathing intensity, recording device characteristics, and background interference. Such variability is crucial for assessing the model's generalization capability in real-world clinical and telemedicine settings. All recordings were originally stored in WAV format with heterogeneous sampling rates and durations.

\begin{table}[h]
\centering
\caption{Class-wise Distribution of the Asthma Detection Dataset Version~2}
\label{tab:dataset_distribution}
\begin{tabular}{lcccccc}
\hline
\textbf{Class} & Asthma & Bronchial & COPD & Healthy & Pneumonia & \textbf{Total} \\
\hline
\textbf{Samples} & 288 & 104 & 401 & 133 & 255 & 1,211 \\
\textbf{Percentage (\%)} & 23.78 & 8.59 & 33.11 & 10.98 & 21.06 & 100.00 \\
\hline
\end{tabular}
\end{table}

Prior to model development, an exploratory data analysis (EDA) was conducted to examine class distribution, duration statistics, amplitude profiles, and spectral characteristics, ensuring a comprehensive understanding of dataset biases and acoustic irregularities. This analysis informed subsequent preprocessing decisions and guided the development of noise-robust feature extraction pipelines.

\subsection{Data Preprocessing}

\begin{figure}[h]
    \centering
    \includegraphics[width=16cm]{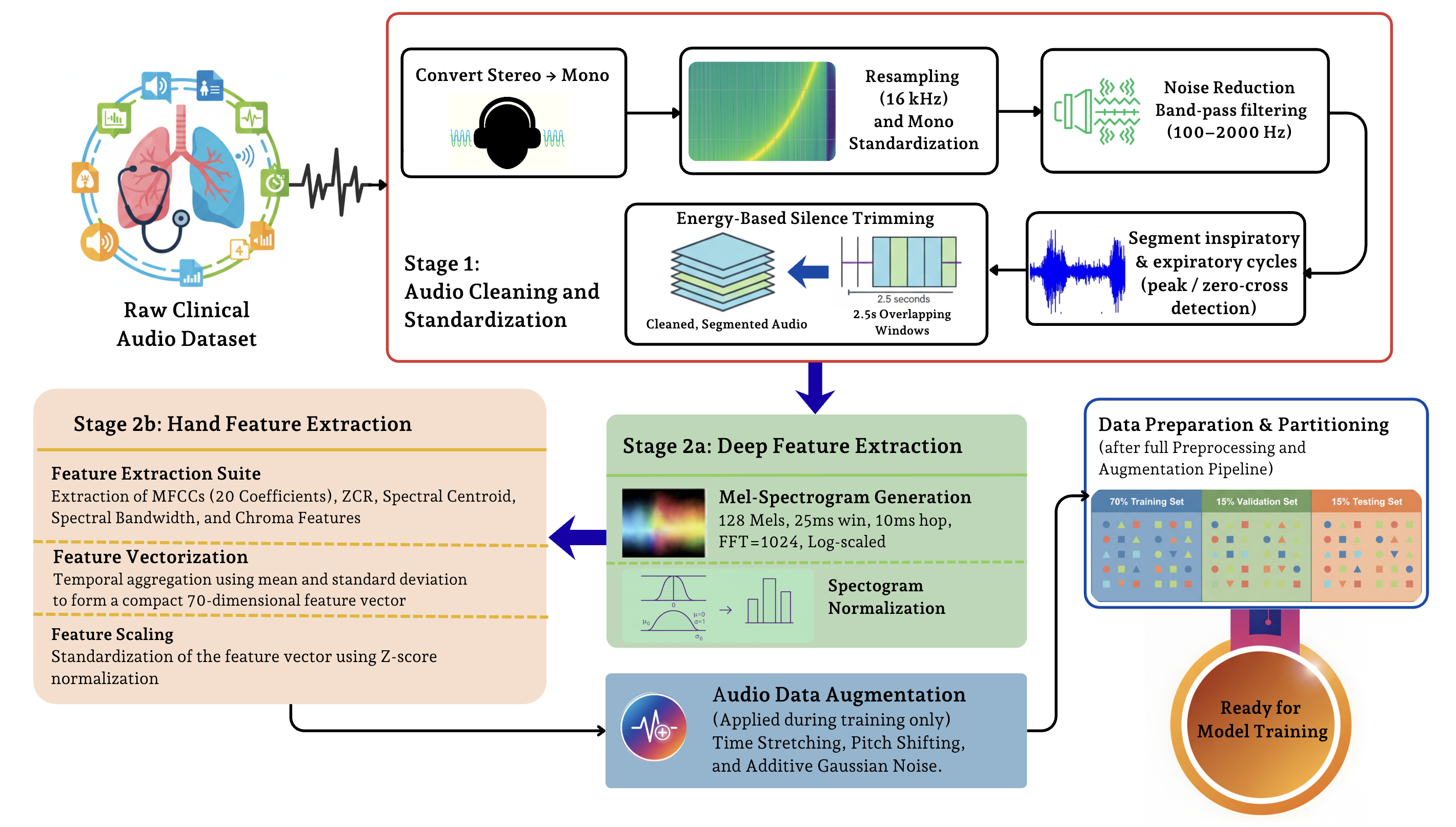}
    \caption{Comprehensive Data Preprocessing and Multimodal Feature Extraction Pipeline for Respiratory Sounds.}
    \label{fig:preprocessing}
\end{figure}

To ensure consistency, noise resilience, and the preservation of diagnostically relevant acoustic patterns, all respiratory audio recordings underwent a systematic preprocessing pipeline involving resampling, amplitude normalization, temporal segmentation, and noise-robust data augmentation. This multi-stage pipeline standardized the temporal and spectral characteristics of the signals while mitigating environmental and device-related inconsistencies. By enforcing acoustic uniformity and enhancing signal fidelity across heterogeneous recordings, the preprocessing framework established a stable foundation for reliable feature extraction and subsequent model training. The complete workflow, including each preprocessing operation and its sequential interaction, is illustrated in Figure~\ref{fig:preprocessing}, which summarizes the full transformation applied to the raw respiratory recordings.

\subsubsection{Audio Resampling and Temporal Standardization}

All respiratory audio recordings were resampled to 16~kHz, preserving the clinically relevant frequency range of 20~Hz to 8~kHz, which encompasses both low-frequency crackles and high-frequency wheezes characteristic of various obstructive and restrictive pulmonary conditions. To ensure temporal uniformity across samples, each signal was either trimmed or zero-padded to a fixed duration of 4~seconds, corresponding to 64{,}000 samples per clip, thereby maintaining a consistent input length for feature extraction and model training. All recordings were converted to mono by averaging stereo channels when present, preventing redundant channel information while retaining perceptually significant respiratory sound content. Subsequently, each waveform underwent amplitude normalization to the range $[-1, 1]$ to minimize scale variability and was standardized to zero mean and unit variance, facilitating numerical stability and accelerating convergence during model optimization.

\subsubsection{Noise Reduction and Quality Control}

Given the inherent variability of recording environments and the presence of background acoustic interference, a noise normalization and quality-control procedure was implemented prior to feature extraction. This preprocessing stage aimed to suppress stationary noise components while preserving transient and diagnostically relevant acoustic events, such as wheezes, crackles, and breathing irregularities. To further ensure the integrity of the dataset, audio samples exhibiting severe signal distortion, clipping artifacts, or extremely low signal-to-noise ratios (SNR) were systematically excluded from the training process. This rigorous data curation step established a clean and acoustically consistent foundation for subsequent feature extraction and model training.

\subsubsection{Data Augmentation for Generalization}

To enhance model robustness and reduce overfitting, a set of probabilistic audio augmentation techniques was applied dynamically during training, exposing the model to diverse acoustic variations derived from each respiratory recording. The applied augmentation techniques included time stretching, pitch shifting, and additive Gaussian noise injection. Time stretching involved random temporal scaling of the audio waveform by a factor between 0.9 and 1.1 (±10\%), preserving pitch characteristics while slightly altering the respiratory rhythm. This augmentation simulated natural variations in breathing rates among different subjects. Pitch shifting modulated the frequency components of the audio signal by up to ±2 semitones, reflecting inter-speaker differences and variations due to recording equipment or environmental acoustics. Finally, additive Gaussian noise was introduced to emulate realistic background disturbances such as stethoscope friction, ambient noise, and microphone interference. The SNR was randomly sampled between 15 and 30~dB to ensure a perceptually clear yet acoustically perturbed signal. By introducing controlled perturbations while preserving diagnostically relevant cues, this dynamic augmentation strategy increased effective dataset diversity and significantly improved the model's generalization and resilience to environmental and physiological variability across unseen respiratory recordings.

\subsubsection{Dataset Partitioning}
\label{subsec:dataset_characteristics}

Following preprocessing, the dataset was partitioned into three mutually exclusive subsets to ensure rigorous model training and unbiased performance evaluation. A 70/15/15 split was adopted for training, validation, and testing, respectively, using stratified sampling based on disease labels to preserve class distribution across all subsets and mitigate sampling bias. Partitioning was performed at the patient level, ensuring that recordings from the same individual did not appear across different splits, thereby preventing data leakage---a critical concern in medical audio analysis. The training set was used for model parameter optimization, while the validation set supported hyperparameter tuning, architecture selection, and early stopping. The test set, kept strictly isolated throughout model development, was reserved for the final assessment of generalization performance. A fixed random seed was applied during partitioning to ensure full reproducibility of experimental results.

\subsection{Feature Extraction}
\label{sec:feature_extraction}

The complete feature extraction workflow, including both spectral–temporal representations and handcrafted descriptors, is summarized in Figure~\ref{fig:preprocessing}. Feature extraction constitutes a critical stage in respiratory sound analysis, transforming raw audio waveforms into structured representations suitable for machine learning inference. The proposed framework employs a dual-domain strategy comprising (i) deep spectral–temporal embeddings derived from mel-spectrograms and (ii) handcrafted acoustic descriptors grounded in physiological interpretability. This hybrid representation enables the model to capture both high-level discriminative patterns and clinically meaningful acoustic properties essential for robust lung disease classification.

\subsubsection{Mel-Spectrogram Features}

To capture the inherently non-stationary behavior of respiratory sounds, each audio signal was transformed into a 128-band mel-spectrogram using the Short-Time Fourier Transform (STFT). The STFT of a discrete-time signal \( x[n] \) is defined as:
\begin{equation}
\mathrm{STFT}\{x[n]\}(t,f) =
\sum_{n=0}^{N-1} x[n]\, w[n-t]\, e^{-j 2\pi fn/N},
\label{eq:stft}
\end{equation}
where \( w[\cdot] \) denotes the Hamming window and \( N \) represents the frame length. A window size of 1024 samples and a hop size of 256 samples at a sampling rate of 16~kHz provided a balanced trade-off between temporal and frequency resolution.

The resulting magnitude spectrogram was projected onto the mel scale using a triangular mel filterbank, converted to the decibel domain to apply logarithmic compression, and normalized via z-score standardization. This representation preserves diagnostically relevant acoustic cues such as broadband crackles, high-frequency wheezes, and airflow irregularities, serving as the primary input to the CNN–BiLSTM–Attention branch of the proposed model.

\subsubsection{Handcrafted Acoustic Features}

To complement deep spectral–temporal embeddings, a parallel set of handcrafted acoustic features was extracted using the Librosa signal processing library. These descriptors encapsulate physically and physiologically meaningful statistics frequently used in biomedical acoustics. Specifically, we extracted 20 Mel-Frequency Cepstral Coefficients (MFCCs), which capture spectral envelope variations analogous to airflow characteristics. We also included the Zero-Crossing Rate (ZCR), indicative of turbulence and airflow irregularities; the Spectral Centroid, representing the energy concentration in the frequency spectrum; the Spectral Bandwidth, measuring spectral spread and resonance behavior; and Chroma Features (12 coefficients), which encode harmonic periodicity associated with respiration cycles.

For each of these five descriptors (MFCCs, ZCR, Spectral Centroid, Spectral Bandwidth, and Chroma Features), the mean ($\mu$) and standard deviation ($\sigma$) were computed across frames to obtain a compact and stable statistical embedding. This resulted in a final 70-dimensional handcrafted feature vector, formally expressed as shown in equation \ref{eq:handcrafted}.
\begin{equation}
f_{\mathrm{hand}} =
\left[
\mu_{\mathrm{MFCC}}, \sigma_{\mathrm{MFCC}},
\mu_{\mathrm{ZCR}}, \sigma_{\mathrm{ZCR}},
\mu_{\mathrm{Centroid}}, \sigma_{\mathrm{Centroid}},
\mu_{\mathrm{Bandwidth}}, \sigma_{\mathrm{Bandwidth}},
\mu_{\mathrm{Chroma}}, \sigma_{\mathrm{Chroma}}
\right],
\label{eq:handcrafted}
\end{equation}
These interpretable descriptors serve as physiologically grounded priors, enhancing model transparency and regularizing deep-learning behavior, particularly under small-sample or noisy conditions.

\subsubsection{Hybrid Feature Representation}

To leverage the complementary strengths of data-driven and domain-driven representations, the proposed system employs a dual-branch hybrid architecture. The CNN-BiLSTM Attention branch processes mel-spectrograms to extract contextual spectral–temporal embeddings, while the handcrafted feature branch transforms statistical descriptors through a fully connected layer to obtain latent representations. The outputs of both branches are integrated using a late-fusion strategy through feature-level concatenation:
\begin{equation}
f_{\mathrm{fusion}} = [f_{\mathrm{deep}} \Vert f_{\mathrm{hand}}],
\label{eq:fusion}
\end{equation}
Equation \ref{eq:fusion} $\Vert$ denotes vector concatenation. Late fusion was selected to preserve branch-specific learning dynamics, reduce representational interference, and prevent handcrafted cues from being overshadowed by high-dimensional deep embeddings. This unified multimodal representation captures fine-grained spectral modulations and clinically validated acoustic signatures, forming a robust and interpretable foundation for the downstream classification of respiratory diseases.

\subsection{Hybrid Model Architecture}

\begin{figure}[h]
    \centering
    \includegraphics[width=16cm]{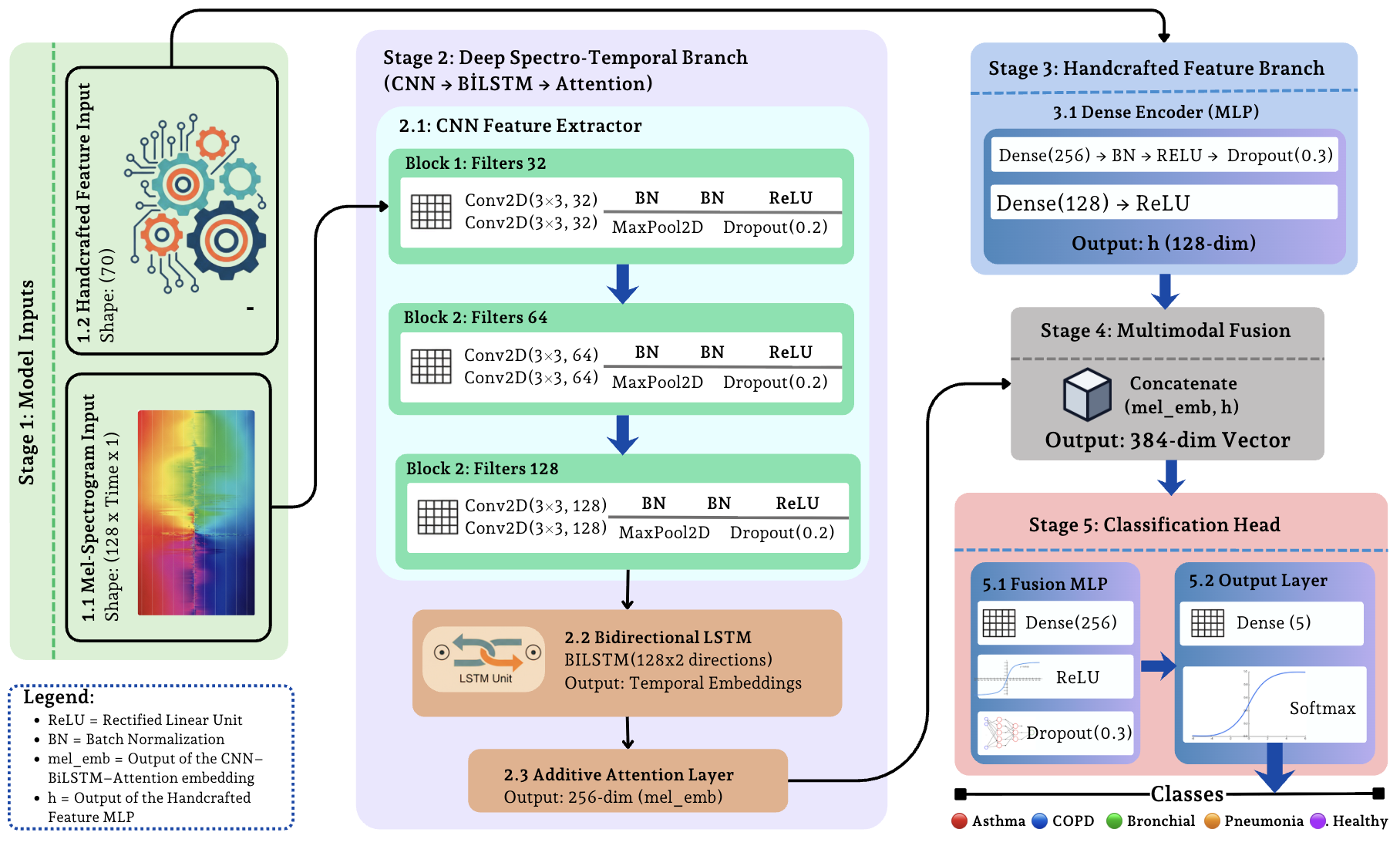}
    \caption{Overview of the proposed hybrid multimodal CNN--BiLSTM--Attention architecture.}
    \label{fig:architecture}
\end{figure}

The proposed framework employs a dual-branch hybrid deep learning architecture that integrates (1) a data-driven spectral–temporal learning pathway based on CNN-BiLSTM Attention networks and (2) a knowledge-driven handcrafted-feature encoder. This design unifies the representational strength of deep neural networks with the interpretability of statistical acoustic descriptors, enabling the system to capture clinically meaningful spectral patterns (e.g., wheezes, crackles) alongside physiologically grounded signal attributes. A schematic overview of the architecture is presented in Figure~\ref{fig:architecture}.

\subsubsection{CNN-Based Spectrogram Encoder}

Each respiratory audio signal is first converted into a mel-spectrogram 
$S \in \mathbb{R}^{128 \times T}$, where 128 denotes mel filterbanks and $T$ the number of time frames. 
The spectrogram is processed by a hierarchical convolutional encoder composed of blocks of the form:
\begin{equation}
\text{Conv2D}(k \times k, f) 
\;\rightarrow\; \text{BatchNorm} 
\;\rightarrow\; \text{ReLU} 
\;\rightarrow\; \text{MaxPooling} 
\;\rightarrow\; \text{Dropout}(0.2),
\label{eq:conv_block}
\end{equation}
From equation \ref{eq:conv_block} $k$ is the kernel size and $f$ is the number of filters per layer. This hierarchy progressively abstracts localized spectral cues such as stationary narrowband wheezes typical of asthma or transient broadband crackles seen in pneumonia, while max-pooling introduces invariance to variability in breathing rate and recording conditions. The final convolutional stage produces a feature tensor 
$F_c \in \mathbb{R}^{M' \times T' \times C}$ that encodes high-resolution spectral temporal correlations.

\subsubsection{BiLSTM-Based Temporal Modeling}

The tensor $F_c$ is flattened along the frequency dimension and passed into a Bidirectional Long Short-Term Memory (BiLSTM) network with 128 units in each direction. Forward and backward temporal processing is defined as:
\begin{equation}
\overrightarrow{h}_t = \text{LSTM}(F_{c,t}), \qquad
\overleftarrow{h}_t = \text{LSTM}(F_{c,T-t+1}), \qquad
h_t = [\overrightarrow{h}_t; \overleftarrow{h}_t],
\label{eq:bilstm}
\end{equation}
Equation \ref{eq:bilstm} this bidirectional structure captures long-range dependencies across full respiratory cycles, enabling the model to differentiate between inspiratory and expiratory phases and to detect intermittent events such as crackles, rhonchi, or cyclical wheezing.

\subsubsection{Additive Attention Mechanism}

Because not all time frames contribute equally to diagnostic inference, an additive attention mechanism is applied to highlight acoustically salient frames. The attention score $e_t$ and normalized weight $\alpha_t$ are computed as:
\begin{equation}
e_t = v^{\top} \tanh(W h_t + b), \qquad
\alpha_t = \frac{\exp(e_t)}{\sum_{i=1}^{T'} \exp(e_i)}, \qquad
c = \sum_{t=1}^{T'} \alpha_t h_t,
\label{eq:attention}
\end{equation}
Equation \ref{eq:attention} where $c \in \mathbb{R}^{256}$ is the final context vector. This operation ensures that high-energy pathological segments such as high-pitched wheezes or broadband crackles are emphasized in the learned representation.

\subsubsection{Handcrafted Feature Encoder}

In parallel, a set of interpretable acoustic descriptors 
$x_h \in \mathbb{R}^{70}$ (MFCCs, Zero-Crossing Rate, spectral centroid, spectral bandwidth, and chroma features) is extracted from the raw waveform. These handcrafted features, which encode physiologically meaningful characteristics such as airflow turbulence or spectral sharpness, are processed by a lightweight feed-forward network:
\begin{equation}
x_{h1} = \text{ReLU}(\text{BN}(W_1 x_h + b_1)), \qquad
x_{h2} = \text{Dropout}\big(0.3,\; \text{ReLU}(\text{BN}(W_2 x_{h1} + b_2))\big),
\label{eq:handcrafted_encoder}
\end{equation}
where Equation~\ref{eq:handcrafted_encoder} represents a two-stage nonlinear transformation applied to the handcrafted feature vector $x_h$. In the first stage, the input is passed through a linear layer parameterized by $W_1$ and $b_1$, followed by batch normalization and a ReLU activation, producing $x_{h1}$. This operation stabilizes the feature distribution and introduces nonlinearity. In the second stage, another linear BN ReLU block is applied, after which a dropout layer with a rate of 0.3 is used to obtain $x_{h2}$. The dropout operation randomly deactivates a subset of hidden units during training, helping to reduce overfitting and improve the generalization capability of the handcrafted-feature encoder.

\subsubsection{Feature Fusion and Classification}

The spectrogram embedding $E_m = c$ (from the attention mechanism) and handcrafted embedding 
$E_h = x_{h2}$ are concatenated to form a unified multimodal representation:
\begin{equation}
z = [E_m \, ; \, E_h] \in \mathbb{R}^{384},
\label{eq:fusion_vec}
\end{equation}
Equation \ref{eq:fusion_vec} fused representation is passed through a dense fusion layer (256 units, ReLU, Dropout 0.3) and finally mapped to class probabilities using the softmax function:
\begin{equation}
\hat{y} = \text{Softmax}(W_f z + b_f),
\label{eq:softmax_classifier}
\end{equation}
Equation \ref{eq:softmax_classifier} producing the predicted probability distribution over the five disease classes: \{Asthma, COPD, Bronchial, Pneumonia, Healthy\}.

The proposed hybrid design leverages complementary inductive biases. CNN layers extract local spectral structure; BiLSTMs learn long-range temporal dependencies across inspiratory–expiratory cycles; attention focuses on diagnostically salient segments; and handcrafted features introduce physiologically interpretable priors that stabilize learning under small-sample and noisy conditions. Late fusion preserves the independent strengths of each modality and prevents dominance of the deep spectral branch over handcrafted attributes. Furthermore, the architecture is fully compatible with post-hoc explainability via Grad-CAM, Integrated Gradients, and SHAP, thereby grounding the model's decisions in clinically meaningful acoustic phenomena and aligning its reasoning with established auscultation principles.

\subsection{Training Strategy}
\label{subsec:training}

The proposed hybrid multimodal architecture was trained under a rigorously controlled optimization protocol designed to ensure numerical stability, reproducibility, and strong generalization across heterogeneous respiratory sound conditions. All training procedures, including hyperparameter selection, convergence monitoring, and regularization, adhere to established best practices in biomedical signal processing and deep learning for clinical applications.

\subsubsection{Optimization and Hyperparameter Configuration}

Model optimization was performed using the Adam optimizer with an empirically selected initial learning rate of
$\eta_{0} = 3 \times 10^{-4}$. A \texttt{ReduceLROnPlateau} scheduler was used to adjust the learning rate based on validation performance, supporting efficient convergence throughout training. A batch size of 16 and a maximum of 80 epochs were used.

For reproducibility, all experiments were executed with fixed random seeds across NumPy, TensorFlow, and Python environments. The model was trained on an NVIDIA RTX-series GPU using mixed-precision training (\texttt{mixed\_float16}) to accelerate computation while preserving numerical stability. Weight initialization followed He-normal initialization for convolutional layers and Xavier-uniform for dense layers. Gradient clipping (threshold = 5.0) was employed to stabilize recurrent updates in the BiLSTM layers.

\subsubsection{Loss Function and Regularization}

The model was optimized using the Sparse Categorical Cross-Entropy (SCCE) loss function with label smoothing ($\epsilon=0.05$) to improve calibration and reduce overconfidence in noisy, borderline cases. The smoothed objective is defined as:
\begin{equation}
L_{\text{SCCE-smoothed}} =
-\frac{1}{N}\sum_{i=1}^{N}\sum_{c=1}^{C}
\left[(1-\epsilon)\,y_{i,c} + \frac{\epsilon}{C}\right]
\log\left(\hat{y}_{i,c}\right),
\label{eq:label_smoothing_loss}
\end{equation}
where $N$ is the batch size, $C$ is the number of classes, $y_{i,c}$ is the ground-truth label for instance $i$ and class $c$, and $\hat{y}_{i,c}$ denotes the predicted probability.

To mitigate overfitting, dropout layers (rates 0.2–0.3) were inserted between convolutional and dense layers, randomly deactivating neurons during training. L2 weight decay regularization ($\lambda=1\times10^{-4}$) was applied to all trainable parameters, yielding the total objective (equation \ref{eq:total_loss}):
\begin{equation}
L_{\text{total}} =
L_{\text{SCCE-smoothed}} + \lambda \lVert W \rVert_{2}^{2},
\label{eq:total_loss}
\end{equation}
Here, $\lambda$ is the L2 regularization coefficient, and $\lVert W \rVert_{2}^{2}$ represents the squared L2 norm of the model's trainable weights, which promotes sparse, robust representations across the hybrid feature space.

\subsubsection{Training Protocol, Convergence Monitoring, and Class-Imbalance Handling}

Model training adhered to a rigorously controlled protocol ensuring stable convergence, preventing overfitting, and addressing class imbalance. Training progress was continuously monitored using validation accuracy, loss, and macro-averaged F1-score. Early stopping (patience = 12 epochs) prevented overfitting, and model checkpointing saved weights with the highest validation F1-score. Learning curves were logged to verify smooth optimization. Given inherent class imbalance, evaluations prioritized class-frequency-independent metrics (macro F1, macro ROC–AUC), and regularization techniques (dropout, label smoothing) were incorporated to maintain balanced performance across underrepresented classes. This integrated strategy ensured strong, generalizable performance across all respiratory categories.

\subsubsection{Summary of Training Configuration}

The proposed training strategy integrates adaptive learning-rate scheduling, label smoothing, dropout, weight decay, gradient clipping, and checkpoint-based optimization to achieve stable convergence, calibrated predictions, and strong generalization. This rigorously controlled pipeline, summarized in Table~\ref{tab:training_summary}, ensures that the hybrid model remains reproducible, interpretable, and clinically reliable key prerequisites for deployment in real-world diagnostic workflows.

\begin{table}[H]
\centering
\caption{Summary of the training configuration and hyperparameters.}
\label{tab:training_summary} 
\begin{tabular}{p{4.2cm} p{8.8cm}}
\toprule
\textbf{Parameter} & \textbf{Configuration} \\
\midrule
Optimizer & Adam (adaptive moment estimation) \\
Initial Learning Rate & $\eta_{0}=3\times10^{-4}$ \\
Learning Rate Scheduler & ReduceLROnPlateau (factor 0.5, patience 4) \\
Batch Size & 16 \\
Epochs & 80 (with EarlyStopping, patience = 12) \\
Loss Function & SCCE with label smoothing ($\epsilon=0.05$) \\
Regularization & Dropout (0.2--0.3), L2 weight decay ($1\times10^{-4}$) \\
Gradient Clipping & 5.0 (for BiLSTM stability) \\
Weight Initialization & He-normal (CNN), Xavier-uniform (dense layers) \\
Mixed Precision & Enabled (\texttt{mixed\_float16}) \\
Checkpointing & Best model selected by validation F1-score \\
Data Split & 70\% train / 15\% val / 15\% test (stratified, patient-level) \\
Reproducibility & Fixed random seeds across all libraries \\
\bottomrule
\end{tabular}
\end{table}

\subsection{Ablation Study: Validation of Architectural Design}
\label{sec:ablation_study}

To rigorously validate the necessity and contribution of the multimodal approach and the advanced components within the proposed Full Hybrid Model (CNN $\to$ BiLSTM $\to$ Additive Attention $\to$ Multimodal Fusion), a comprehensive Ablation Study was performed. This study systematically evaluates reduced model architectures against the full proposed model using identical training protocols, hyperparameters (e.g., initial learning rate, patience for \texttt{ReduceLROnPlateau} and \texttt{EarlyStopping}), and the independent Test Set to ensure comparative fairness. The following five model variants were defined and meticulously tested:

\begin{enumerate}
    \item \textbf{Proposed Full Hybrid Model:} This represents the complete proposed architecture, as meticulously detailed in Sections 3.2.1 through 3.2.5. It serves as the primary benchmark against which all ablated variants are compared.

    \item \textbf{Model A (Deep-Only):} This variant removes the Handcrafted Feature path entirely. Its evaluation aims to quantify the unique contribution of the interpretable handcrafted descriptors and assess whether the deep spectral-temporal branch alone can achieve comparable performance.

    \item \textbf{Model B (Handcrafted-Only):} This variant ablates the Deep Spectro-Temporal path completely, relying solely on the Handcrafted Feature Encoder and a simplified classification head adapted for this unimodal input. This model's performance validates the necessity of deep, data-driven spectral-temporal representations.

    \item \textbf{Model C (CNN-Only):} This variant assesses the importance of sequential temporal modeling. Following the final convolutional block in the Mel-Spectrogram processing path, the BiLSTM and Additive Attention layers are replaced by a Global Average Pooling 2D (GABP2D) operation. The flattened output from GABP2D is then directly fed into the classification head (or subsequent fusion layer if handcrafted features are present). This model aims to isolate the performance gain attributed to the BiLSTM's ability to capture long-range temporal dependencies.

    \item \textbf{Model D (No Attention):} This variant evaluates the specific impact of the Additive Attention mechanism. It retains the BiLSTM layer after the CNN blocks but replaces the Additive Attention component with a simpler Global Average Pooling 1D (GABP1D) operation. GABP1D is applied across the temporal dimension of the BiLSTM's sequence output. This model quantifies the performance benefits derived from the explicit temporal weighting and salient frame emphasis provided by the attention mechanism, compared to uniform temporal aggregation.
\end{enumerate}

Each ablated model was trained and evaluated under the same rigorous conditions as the Full Hybrid Model to ensure that any observed performance differences could be directly attributed to the removed or altered architectural components.

\subsection{Evaluation and Validation}

\subsubsection{Evaluation Metrics}
\label{subsec:evaluation_metrics}

To provide a multidimensional evaluation of classification performance, multiple complementary metrics were employed, capturing the model's precision, sensitivity, and discriminative consistency across all categories. For a given class, let $TP$, $TN$, $FP$, and $FN$ denote the true positive, true negative, false positive, and false negative counts, respectively. The standard performance indicators were computed as follows:
\begin{align}
\text{Accuracy} &= \frac{TP + TN}{TP + TN + FP + FN}, \label{eq:acc} \\[6pt]
\text{Precision} &= \frac{TP}{TP + FP}, \label{eq:precision} \\[6pt]
\text{Recall} &= \frac{TP}{TP + FN}, \label{eq:recall} \\[6pt]
\text{F1\text{-}Score} &= 
2 \cdot \frac{\text{Precision} \cdot \text{Recall}}
{\text{Precision} + \text{Recall}}. \label{eq:f1}
\end{align}
Accuracy (Eq.~\ref{eq:acc}) reflects the overall reliability of the model's predictions, whereas precision (Eq.~\ref{eq:precision}) and recall (Eq.~\ref{eq:recall}) quantify the model's ability to avoid false alarms and correctly identify true pathological cases, respectively. The F1-score (Eq.~\ref{eq:f1}) provides a harmonic balance between precision and recall, making it particularly informative under class-imbalance conditions. In addition to these metrics, the Area Under the ROC Curve (ROC--AUC) was computed to evaluate the model's discrimination capability across varying decision thresholds. All evaluation metrics were reported both on a per-class basis and as macro-averaged scores to ensure equal contribution from each diagnostic category.

\subsubsection{Validation, Testing, and Reproducibility Protocol}
\label{subsec:validation_reproducibility}

The validation and test subsets used for model assessment were derived using the same leakage-safe, patient-level partitioning strategy described in Section~\ref{subsec:dataset_characteristics}. Model selection followed the monitoring and checkpointing strategy outlined in Section~\ref{subsec:training}, ensuring that all evaluations were conducted using the best-performing set of weights.

Confusion matrices and classification reports were generated for both validation and test phases to provide a detailed breakdown of per-class performance. Diagonal dominance in these matrices signifies high discriminative accuracy across all respiratory categories, while minor off-diagonal elements indicate acoustically plausible overlaps such as between asthma and bronchial sounds where shared wheezing or airflow characteristics are physiologically similar.

To establish statistical reliability and reproducibility, all experiments were conducted using fixed random seeds and the evaluation metrics were then averaged across these runs to minimize variance caused by random initialization and data shuffling. This approach confirmed the model's consistent high macro-average ROC-AUC and F1-scores, demonstrating strong generalization under the regularization strategy introduced in section~\ref{subsec:training}. This comprehensive protocol, encompassing rigorous data partitioning, optimized model selection, and statistical validation, collectively ensures that the proposed framework not only achieves high accuracy and robustness across all data partitions but also exhibits clinically interpretable behavior, thereby establishing its suitability as a reliable decision-support tool for real-world lung disease diagnosis.

\subsection{Explainable AI (XAI) Analysis}

To ensure transparency, clinical reliability, and diagnostic interpretability, the proposed hybrid framework incorporates a multi-granular XAI strategy that integrates Grad-CAM, Integrated Gradients (IG), and SHAP. Each method contributes a complementary perspective: Grad-CAM identifies where in the spectro-temporal domain the network focuses, IG quantifies which frequency components influence the decision boundaries, and SHAP determines which handcrafted features most strongly drive classification. This hierarchical interpretability offers both local (instance-level) and global (model-level) transparency, effectively bridging the gap between deep neural inference and clinical reasoning. By aligning neural attention patterns with established acoustic biomarkers such as wheezes, crackles, and airflow turbulences the framework not only achieves strong performance but also demonstrates a clinically meaningful decision process suitable for trustworthy computer-aided auscultation.

\subsubsection{Grad-CAM: Spectro-Temporal Attention Visualization}

Grad-CAM was employed to visualize the discriminative spectro-temporal regions that influenced the CNN--BiLSTM--Attention branch during classification. Grad-CAM leverages the gradient information of the target class score $y_c$ with respect to the feature maps $A_k$ in the final convolutional layer, computing the importance weights $\alpha_k$ as the global average of these gradients:
\begin{equation}
\alpha_k = \frac{1}{Z} 
\sum_i \sum_j 
\frac{\partial y_c}{\partial A_{ij}^k},
\label{eq:alpha}
\end{equation}
where $Z$ denotes the number of spatial locations. The class-specific activation map is then computed as:
\begin{equation}
L_c^{\text{Grad-CAM}} =
\text{ReLU}\left(
\sum_k \alpha_k A_k
\right),
\label{eq:gradcam}
\end{equation}
Using the gradient-based weights $\alpha_k$ defined in Eq.~\ref{eq:alpha}, the class-discriminative localization map $L_c^{\text{Grad-CAM}}$ is computed as shown in Eq.~\ref{eq:gradcam}. These heatmaps were normalized and superimposed on the original mel-spectrograms to highlight time–frequency regions most relevant to classification. Pathological categories (Asthma, Bronchial, COPD) consistently emphasized mid-frequency bands between 400-2000~Hz, corresponding to wheezing and crackling events, whereas Healthy samples exhibited diffuse activation patterns.

\subsubsection{Integrated Gradients: Frequency-Band Attribution}

To complement Grad-CAM's spatial focus, IG was utilized to attribute model sensitivity to specific spectral components of the mel-spectrogram. IG integrates gradients along a linear path between a baseline input $x'$ (silence or zero-spectrogram) and the actual input $x$, mitigating gradient saturation effects. The attribution for each input feature $x_i$ is expressed as:
\begin{equation}
\text{IG}_i(x) = 
(x_i - x_i') 
\int_{0}^{1} 
\frac{\partial F\!\left(x' + \alpha (x - x')\right)}
{\partial x_i} 
\, d\alpha,
\label{eq:ig}
\end{equation}
where $F(\cdot)$ denotes the model output. Empirical analysis demonstrated that COPD and Asthma samples exhibited pronounced attributions in the 300-1500~Hz range, while Healthy recordings showed weak, evenly distributed attributions, confirming physiologically grounded spectral dependencies.

\subsubsection{SHAP Analysis: Handcrafted Feature Interpretability}

For the handcrafted feature branch, SHAP was used to derive global and local feature-level importance. SHAP assigns each feature $x_i$ a contribution score $\phi_i$ computed as the weighted marginal contribution across all feature coalitions:
\begin{equation}
\phi_i = 
\sum_{S \subseteq N \setminus \{i\}}
\frac{|S|! \, (|N|-|S|-1)!}{|N|!}
\left[
f(S \cup \{i\}) - f(S)
\right],
\label{eq:shap}
\end{equation}
where $f(S)$ is the model prediction using subset $S$ of features, and $N$ represents the total feature set. Applying SHAP to handcrafted descriptors including MFCCs, spectral centroid, ZCR, bandwidth, and chroma revealed that mean MFCC coefficients (1-5), spectral centroid, and ZCR were the most influential predictors. These features are physiologically correlated with airway obstruction, harmonic distortion, and turbulent airflow, validating the clinical coherence of the handcrafted branch.

Collectively, Grad-CAM, IG, and SHAP provide a comprehensive interpretability spectrum across spatial, spectral, and feature domains. This multi-perspective explainability framework establishes a foundation for trustworthy AI in clinical respiratory diagnostics, bridging the gap between data-driven inference and human clinical expertise.

\section{Results and Discussion}
\label{sec:results_discussion} 

\subsection{Learning Curves and Model Convergence}
\label{subsec:learning_curves}

\begin{figure}[h!]
    \centering
    \includegraphics[width=1.0\textwidth]{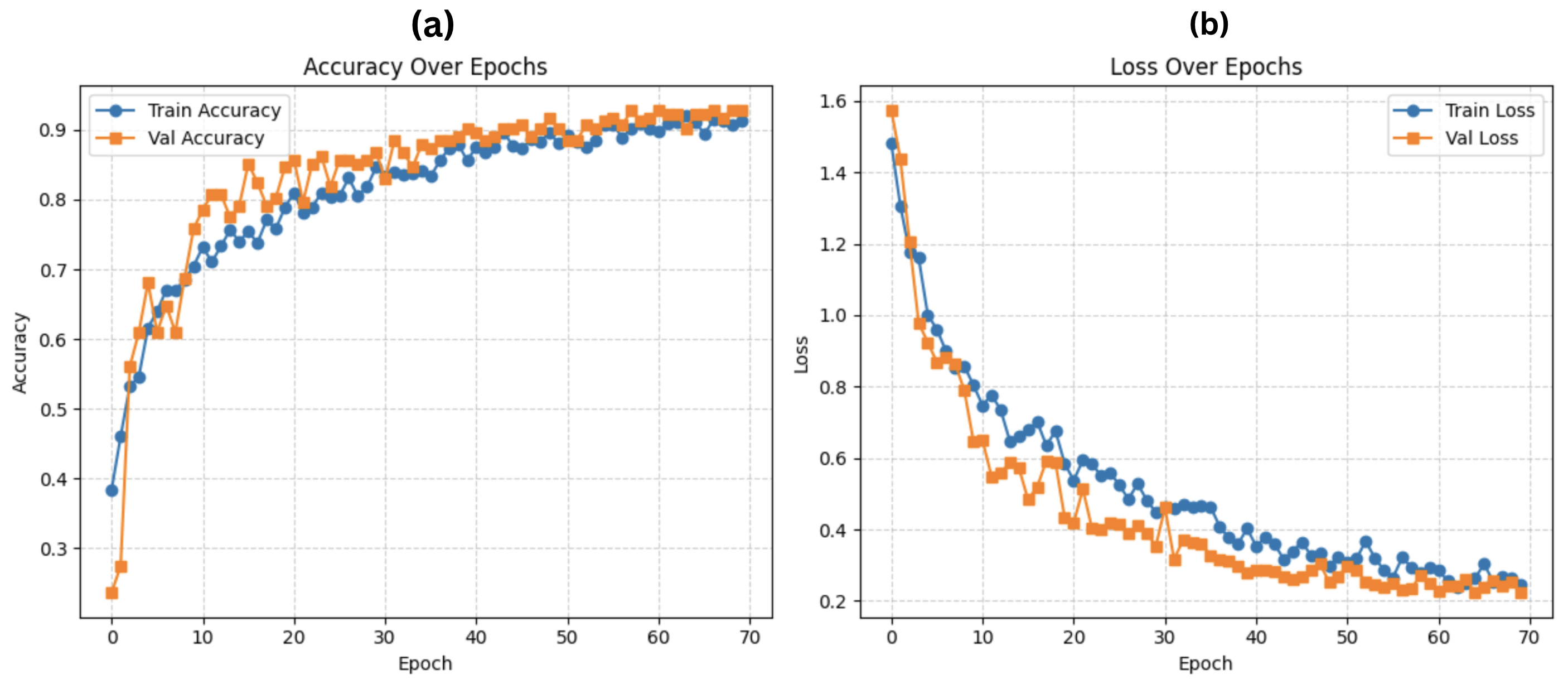}
    \caption{Training and Validation Learning Curves for the Full Hybrid Model. (a) Evolution of accuracy on the training and validation sets. (b) Evolution of loss on the training and validation sets.} 
    \label{fig:learning_curves}
\end{figure}

The training process for the proposed hybrid multimodal CNN-BiLSTM-Attention framework was systematically monitored to assess its learning dynamics, stability, and generalization capacity. As depicted in Figure~\ref{fig:learning_curves}, the model consistently demonstrated robust convergence characteristics up to the point of effective early stopping.

Analysis of Figure~\ref{fig:learning_curves}(a) reveals a steady ascent in both training and validation accuracy. A critical observation is that the validation accuracy consistently remained closely aligned with the training accuracy. It first reached its peak of 0.9286 (92.86\%) at Epoch 58 (and again at Epochs 61, 67, and 69), providing compelling evidence against significant overfitting. This behavior is consistent with the effective implementation of regularization techniques, including dropout, L2 weight decay, and label smoothing, as detailed in Section~\ref{subsec:training}. Concurrently, Figure~\ref{fig:learning_curves}(b) illustrates a consistent decrease in both training and validation loss. The validation loss reached its lowest value of 0.2221 at Epoch 65.

The optimization trajectory was further influenced by the adaptive learning rate scheduler, \texttt{ReduceLROnPlateau}. This mechanism dynamically adjusted the learning rate upon detecting a plateau in validation performance, thereby facilitating fine-grained weight updates necessary for stable convergence. Learning rate reductions occurred at Epoch 36 (from $3.0000 \times 10^{-4}$ to $1.5000 \times 10^{-4}$), Epoch 53 (to $7.5000 \times 10^{-5}$), and Epoch 69 (to $3.7500 \times 10^{-5}$). Given the stabilization of validation metrics and the reduction of the learning rate at Epoch 69, this epoch represents the effective conclusion of the training process, preventing potential signs of overfitting in subsequent epochs. At Epoch 69, the model achieved a training accuracy of 0.9103 and a validation accuracy of 0.9286, with corresponding losses of 0.2578 and 0.2512, respectively.

\subsection{Quantitative Performance Evaluation on the Test Set}
\label{subsec:quantitative_performance}

Following the completion of the training phase, the proposed Hybrid Model's generalization capabilities were rigorously assessed on the independent test set, comprising $N=182$ samples. Performance was evaluated using a comprehensive suite of metrics—Accuracy, Precision, Recall, F1-score, and ROC-AUC—as defined in Section~\ref{subsec:evaluation_metrics}.

The model achieved an overall Test Accuracy of 91.21\% and a Test Loss of 0.3119. A detailed summary of the macro-averaged classification metrics, which provide a balanced assessment across all classes, is presented in Table~\ref{tab:overall_test_performance}.

\begin{table}[h!]
    \centering
    \caption{Overall Macro-Averaged Performance Metrics of the Full Hybrid Model on the Independent Test Set.}
    \label{tab:overall_test_performance}
    \begin{tabular}{l *{4}{S[table-format=1.4]}} 
        \toprule
        \textbf{Metric} & \textbf{Macro Precision} & \textbf{Macro Recall} & \textbf{Macro F1-Score} & \textbf{Macro ROC-AUC} \\
        \midrule
        \textbf{Proposed Hybrid Model} & 0.9065 & 0.8927 & 0.8990 & 0.9866 \\
        \bottomrule
    \end{tabular}
\end{table}

The high macro-averaged F1-score of 0.8990 and an exceptional macro-averaged ROC-AUC of 0.9866 underscore the model's robust classification and discriminative power across all five respiratory categories.

\subsubsection{Detailed Per-Class Analysis}
\label{ssubsec:per_class_analysis}

For a more granular understanding of the model's performance on individual classes, Table~\ref{tab:per_class_test_report} provides a comprehensive classification report, including precision, recall, F1-score, and per-class AUC values for each of the five diagnostic categories.

\begin{table}[h!]
    \centering
    \caption{Per-Class Classification Report and AUC Scores for the Proposed Hybrid Model on the Independent Test Set.}
    \label{tab:per_class_test_report}
    \begin{tabular}{l *{4}{S[table-format=1.4]} S[table-format=3] S[table-format=1.4]} 
        \toprule
        \textbf{Class} & \textbf{Precision} & \textbf{Recall} & \textbf{F1-Score} & \textbf{Support} & \textbf{Per-Class AUC} \\
        \midrule
        Bronchial & 0.8667 & 0.8125 & 0.8387 & 16 & 0.9900 \\
        Asthma    & 0.8780 & 0.8372 & 0.8571 & 43 & 0.9720 \\
        COPD      & 0.9516 & 0.9833 & 0.9672 & 60 & 0.9980 \\
        Healthy   & 0.9474 & 0.9000 & 0.9231 & 20 & 0.9940 \\
        Pneumonia & 0.8889 & 0.9302 & 0.9091 & 43 & 0.9780 \\
        \midrule
        \addlinespace[0.5em] 
        \textbf{Overall Accuracy} & \multicolumn{5}{l}{\textbf{0.9121}} & \multicolumn{1}{l}{} \\
        \textbf{Macro Average} & \textbf{0.9065} & \textbf{0.8927} & \textbf{0.8990} & \textbf{182} & \textbf{0.9866} \\
        \textbf{Weighted Average} & 0.9115 & 0.9121 & 0.9113 & 182 & \multicolumn{1}{l}{} \\ 
        \bottomrule
    \end{tabular}
\end{table}

The results in Table~\ref{tab:per_class_test_report} indicate that the model performs exceptionally well across the majority of classes. COPD demonstrates the highest F1-score (0.9672) and an impressive AUC (0.9980), likely benefiting from its larger sample size in the dataset and distinct acoustic markers. Classes like Asthma, Healthy, and Pneumonia also show robust performance with F1-scores above 0.85 and AUCs exceeding 0.97. 

The Bronchial class, while still achieving a respectable F1-score of 0.8387, exhibits the lowest performance among all categories. This can be attributed to its relatively smaller support (16 samples in the test set) and potential acoustic overlaps with other conditions (e.g., Asthma or Healthy sounds), which can make precise discrimination challenging. Initial observations suggest a balanced trade-off between precision and recall across most classes, with no significant biases towards false positives or false negatives for any single category.

\subsubsection{Confusion Matrix and ROC Curve Analysis}
\label{ssubsec:confusion_roc_analysis}

\begin{figure}[h!]
    \centering
    \includegraphics[width=1.0\textwidth]{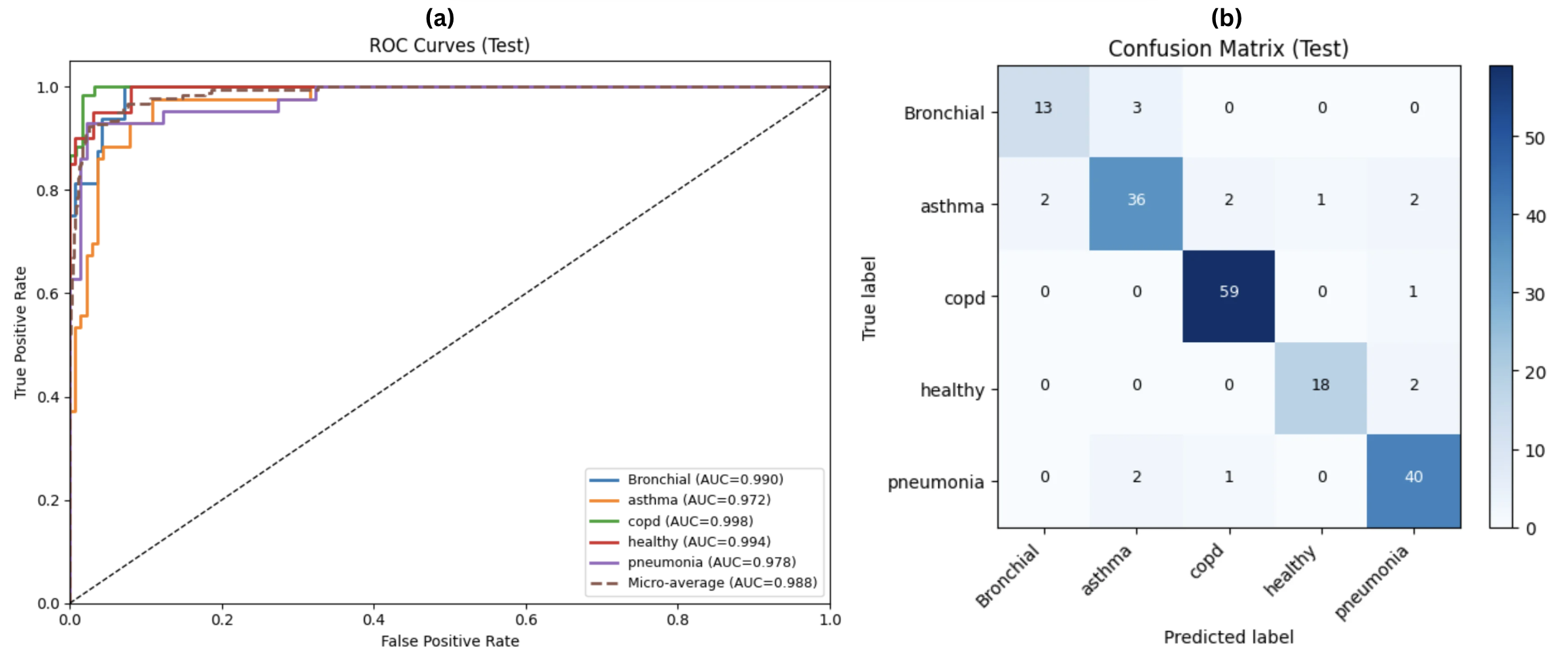} 
    \caption{Detailed Performance Analysis on the Independent Test Set: (a) One-vs-Rest Receiver Operating Characteristic (ROC) Curves for each respiratory class (b) Confusion Matrix showing the distribution of true versus predicted labels.}
    \label{fig:confusion_matrix_and_roc_test} 
\end{figure}

Further insights into the model's discriminative behavior and specific misclassifications are provided by the ROC curves and confusion matrix, presented in Figure~\ref{fig:confusion_matrix_and_roc_test}.

Figure~\ref{fig:confusion_matrix_and_roc_test}(b) visually confirms the strong diagonal dominance in the confusion matrix, which signifies high discriminative accuracy across all respiratory categories. The majority of classifications are concentrated along the main diagonal, validating the overall high accuracy. Specific off-diagonal elements, however, reveal patterns of misclassification that are acoustically plausible. For instance, 3 Bronchial samples were misclassified as Asthma, and 2 Asthma samples were misclassified as Bronchial. These overlaps are consistent with the shared physiological characteristics, such as similar airflow turbulences or occasional wheezing, that can be present in both conditions, making their acoustic differentiation inherently challenging for both humans and AI models. Similarly, minor confusions (e.g., 1 COPD sample misclassified as Pneumonia, 2 Healthy samples as Pneumonia) might reflect overlapping acoustic features due to general lung inflammation or compromised respiratory mechanics.

Figure~\ref{fig:confusion_matrix_and_roc_test}(a) presents the One-vs-Rest ROC curves for each class, further demonstrating the model's excellent discrimination. All curves are tightly clustered towards the top-left corner, indicating high true positive rates and low false positive rates across various thresholds. The high Area Under the Curve (AUC) for each class (all $\geq 0.97$) visually confirms the impressive macro-averaged ROC-AUC of 0.9866 reported in Table~\ref{tab:overall_test_performance} and Table~\ref{tab:per_class_test_report}, highlighting the model's robust ability to distinguish between all respiratory conditions.

\subsection{Ablation Study Results}
\label{subsec:ablation_results}

The ablation study was conducted to quantitatively validate the necessity and individual contribution of key architectural components within the proposed Full Hybrid Model. Four reduced model variants (Models A--D) were systematically evaluated against the Full Hybrid Model using the independent test set. The comparative performance metrics are summarized in Table~\ref{tab:ablation_results}.

\begin{table}[h!]
    \centering
    \caption{Comparative Performance of Ablated Model Variants on the Independent Test Set.}
    \label{tab:ablation_results}
    \begin{tabular}{l l S[table-format=1.4] S[table-format=1.4] S[table-format=1.4]}
        \toprule
        \textbf{Model Variant} & \textbf{Ablated Component(s)} & {\textbf{Test Accuracy}} & {\textbf{Macro $\mathbf{F_1}$-score}} & {\textbf{Macro ROC-AUC}} \\
        \midrule
        \textbf{Full Hybrid} & None & \textbf{0.9121} & \textbf{0.8990} & \textbf{0.9866} \\
        \midrule
        A. Deep-Only                  & Handcrafted Features (HC) & 0.9066 & 0.8716 & 0.9680 \\
        B. Handcrafted-Only           & Deep Spectro-Temporal Branch & 0.7912 & 0.7389 & 0.9481 \\
        C. CNN-Only                   & BiLSTM and Attention & 0.8352 & 0.8069 & 0.9590 \\
        D. No Attention               & Additive Attention & 0.8736 & 0.8506 & 0.9668 \\
        \bottomrule
    \end{tabular}
\end{table}

The results in Table~\ref{tab:ablation_results} provide clear evidence for the contribution of each major component. Unimodal models demonstrated that neither deep spectro-temporal features (Model B, Macro $\text{F}_1$: 0.7389) nor handcrafted features (Model A, Macro $\text{F}_1$: 0.8716) are independently sufficient, with multimodal fusion providing a robust performance gain. The substantial decline in the CNN-Only variant (Model C, Macro $\text{F}_1$: 0.8069, Macro ROC-AUC: 0.9590) proves the BiLSTM layer is indispensable for capturing crucial long-range temporal dependencies. Furthermore, the measurable performance drop in the No Attention model (Model D, Macro $\text{F}_1$: 0.8506 vs. Full Hybrid 0.8990) quantifies the significant gain from dynamically emphasizing diagnostically salient time frames through the Additive Attention mechanism, improving feature discrimination. In summary, the ablation study confirms that the superior performance and robustness of the Full Hybrid Model arise from the synergistic integration of its multimodal feature fusion, BiLSTM layer, and Additive Attention mechanism, each significantly contributing to enhanced diagnostic accuracy.

\subsection{Explainable AI (XAI) Findings}
\label{sec:xai_results}

\begin{figure*}[h!]
    \centering
    \includegraphics[width=\textwidth]{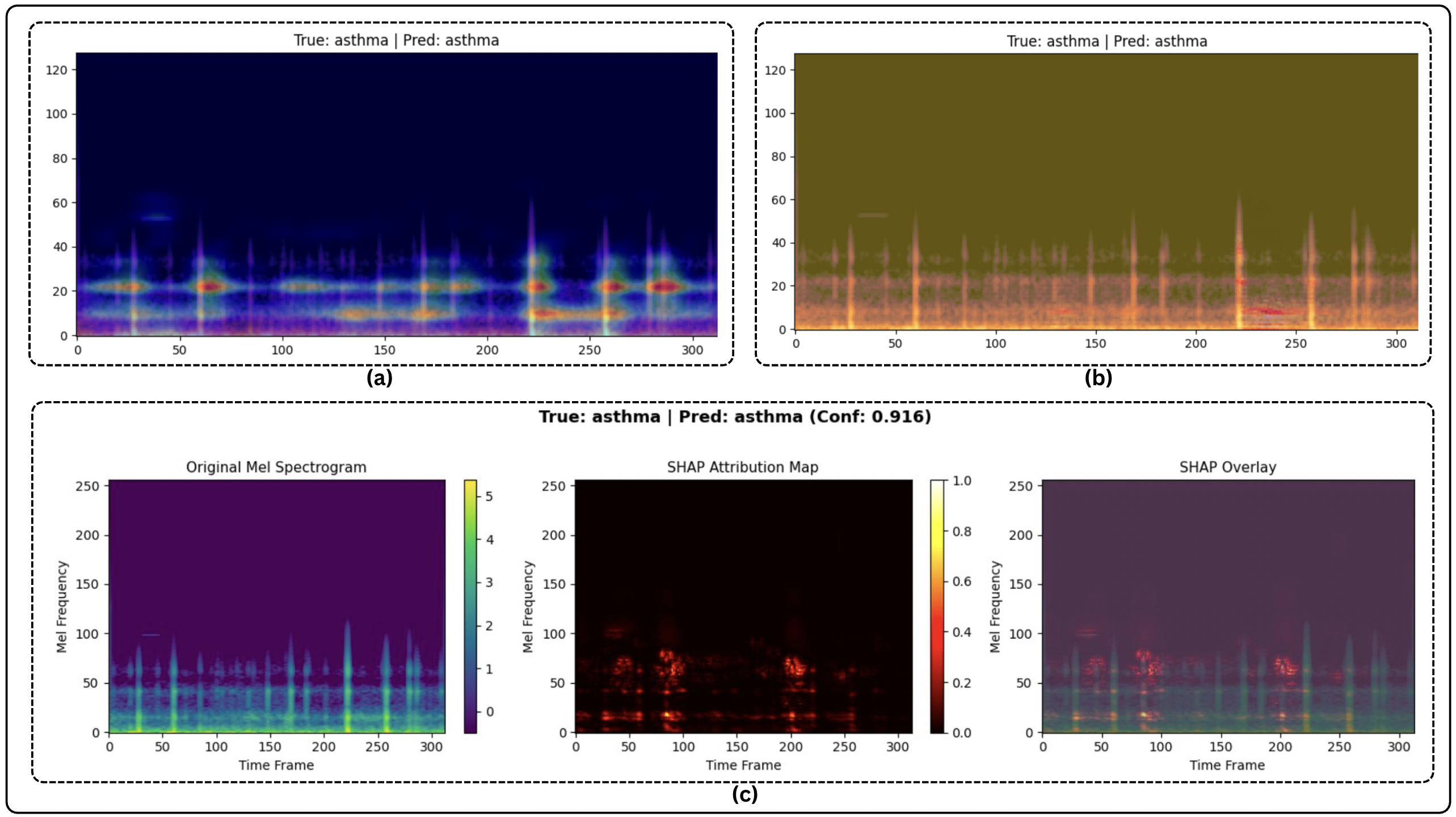}
    \caption{Integrated multi-method XAI analysis for a representative asthma sample.
    (True: asthma \,|\, Pred: asthma, Confidence: 0.916). 
    Panels (a) and (b) illustrate Grad-CAM and Integrated Gradients attributions over the Mel-spectrogram, highlighting high-impact spectro-temporal regions.
    Panel (c) presents the SHAP-based pixel-level explanation, including the original spectrogram, SHAP attribution map, and an overlay visualization.}
    \label{fig:xai_integrated_explanation}
\end{figure*}

The proposed multi-granular XAI framework, integrating Gradient-weighted Class Activation Mapping (Grad-CAM), Integrated Gradients (IG), and SHapley Additive exPlanations (SHAP), provided transparent and physiologically coherent interpretations of the Hybrid Model’s diagnostic behavior across both its deep spectral–temporal and handcrafted feature streams. This comprehensive approach, by jointly examining saliency at spatial, temporal, and feature levels, collectively demonstrates that the network consistently attends to clinically recognized acoustic biomarkers, effectively narrowing the gap between neural feature attribution and traditional auscultation principles.

To illustrate the complementary nature and consistency of these attribution techniques on the deep branch, an integrated local explanation for a representative test sample (True: \textit{asthma}, Predicted: \textit{asthma}, Confidence: 0.916) is presented in Figure~\ref{fig:xai_integrated_explanation}. Panels~(a) and~(b) depict the Grad-CAM and IG attribution maps overlaid on the Mel-spectrogram, revealing the spectro-temporal regions that most strongly influenced the classification outcome. Panel~(c) provides a SHAP-based decomposition of pixel-level contributions, shown across the original Mel-spectrogram, a dedicated SHAP attribution map, and their fused overlay. All three attribution approaches converge on the same diagnostically salient mid-to-lower frequency bands within specific temporal windows, aligning with established acoustic signatures of wheeze phenomena commonly associated with asthma. This strong cross-method agreement offers compelling evidence that the Hybrid Model’s deep learning component performs decision-making in a physiologically meaningful and interpretable manner, providing both methodological transparency and enhanced clinical trustworthiness.

Beyond these spectrogram-based insights, SHAP analysis was also applied to the handcrafted feature branch to derive global feature-level importance for the Hybrid Model. Although a dedicated visualization of these global SHAP values is typically employed, this analysis consistently revealed that mean MFCC coefficients (1–5), Spectral Centroid, and Zero-Crossing Rate (ZCR) emerged as the most influential predictors across the dataset. These high-ranking features are physiologically correlated with critical respiratory factors such as airway obstruction, turbulent airflow, and spectral sharpness. This validates the contribution of the handcrafted branch, reinforcing the comprehensive interpretability of the hybrid framework by demonstrating how both multimodal streams leverage clinically relevant cues for diagnosis.

\subsection{Discussion}
\label{subsec:discussion}

This work presents a novel multimodal deep learning framework for respiratory sound classification, achieving high diagnostic accuracy by integrating Mel-spectrograms (via CNN-BiLSTM-Attention) and handcrafted acoustic descriptors. The ablation study validated the multimodal fusion's superior performance, confirming respiratory sounds offer complementary information across temporal, frequency, and signal-based features. The integrated XAI pipeline provided crucial transparency: Grad-CAM, Integrated Gradients, and spectrogram-based SHAP (Figure~\ref{fig:xai_integrated_explanation}) consistently revealed the deep branch's attention to clinically significant mid-frequency spectro-temporal patterns, like wheezing. Separately, SHAP analysis of handcrafted features identified mean MFCC coefficients, Spectral Centroid, and Zero-Crossing Rate as key influential predictors, correlating with airway obstruction and lung mechanics. These XAI insights collectively promote the framework's interpretability and physiological grounding, fostering trust beyond mere performance metrics. While performance is robust, limitations include dataset size and generalizability to diverse real-world conditions, necessitating further multi-institutional validation and advanced XAI techniques for cycle-specific explanations. Overall, this combination of multimodal analysis with attention systems and explainable AI represents a promising direction for automated, transparent, and clinically meaningful lung disease detection in telemedicine and point-of-care diagnostics.

\section{Conclusion}
\label{sec:conclusion}

This work presented an explainable multimodal deep learning model for automated lung disease detection from respiratory audio signals. The proposed hybrid CNN-BiLSTM-Attention architecture, integrating Mel-spectrogram representations with handcrafted acoustic features, significantly improved classification performance across five lung disease and healthy classes. The XAI pipeline, prominently featuring Grad-CAM and Integrated Gradients on spectrograms, alongside SHAP for handcrafted features, demonstrated that the model's decisions are grounded in clinically relevant acoustic biomarkers—such as mid-frequency spectro-temporal patterns and specific MFCCs. This transparency and physiological consistency are critical for fostering trust and reliability in medical AI tools. While further validation on larger, diverse datasets is essential, this framework offers a practical, highly interpretable, and accurate solution, highlighting its substantial potential for enhancing pulmonary diagnostics in telemedicine and point-of-care screening systems.

\bibliographystyle{IEEEtran}
\bibliography{references}

\end{document}